\documentclass[12pt,twoside]{amsart}

\usepackage{amsfonts,amsthm,amssymb,amsmath,latexsym}
\usepackage{rotating}
\usepackage{booktabs}
\usepackage{multirow}
\usepackage{url}

\title[Computational Modelling with Matlab, Octave and Scilab]{How good are MatLab, Octave and Scilab for Computational Modelling?}
\vspace{5pt}

\thanks{The authors receive support from CNPq, Capes and Fapeal.}

\author[E.\ S.\ Almeida et al.]{\tiny Eliana S.\ de Almeida\and Antonio C.\ Medeiros\and Alejandro C.\ Frery}
\date{}

\begin{document}
\maketitle

\vspace{-20pt}
\begin{center}
{\tiny Laborat\'orio de Computa\c c\~ao Cient\'ifica e An\'alise Num\'erica (LaCCAN)\\
Centro de Pesquisas em Matem\'atica Computacional (CPMAT),\\
Universidade Federal de Alagoas\\
BR 104 Norte km 97, 57072-970 Macei\'o, AL -- Brazil \\
E-mails: acfrery@gmail.com / eliana.almeida@gmail.com / medeiros.tonny@gmail.com  
}\end{center}

\hrule

\begin{abstract}
In this article we test the accuracy of three platforms used in computational modelling: MatLab,  Octave and Scilab, running on i386 architecture and three operating systems (Windows, Ubuntu and Mac OS).
We submitted them to numerical tests using standard data sets and using the functions provided by each platform. 
A Monte Carlo study was conducted in some of the datasets in order to verify the stability of the results with respect to small departures from the original input.
We propose a set of operations which include the computation of matrix determinants and eigenvalues, whose results are known. 
We also used data provided by NIST (National Institute of Standards and Technology), a protocol which includes the computation of basic univariate statistics (mean, standard deviation and first-lag correlation), linear regression and extremes of probability distributions. 
The assessment was made comparing the results computed by the platforms with certified values, that is, known results, computing the number of correct significant digits. 
\end{abstract}

\medskip
\noindent

\medskip
\noindent
\medskip

\hrule

\section{Introduction}\label{sec:1}

Mathematical modelling aims at solving complex problems which can be described in a rigorous mathematical way that enables the use of computers for finding the solution.

Many mathematical models which arise in diverse areas as engineering, bioinformatics and ecology rely on partial differential equations (PDE) or ordinary differential equations (ODE), where the high number of variables requires strong computational effort in their solution.

The final output is the result of a myriad of often disregarded intermediate computations.
To illustrate this, when a finite element mesh is used to perform a structural analysis, computing matrix inversions and determinants are important commonplace operations which are rarely checked.

The search for best approximate solutions, considering some reasonable bounds of errors, imposes tight accuracy requirements on the computational platforms and its libraries or functions. 
When dealing with huge structures described by irregular meshes, for instance, algorithms for domain partitioning and parallel computing are often needed, and the correctness of the results is still more critical. 
Such partitioning algorithms are usually based on either topological or spectral methods, which assess the algebraic properties of the graph associated to the mesh~\cite{Nasra90, Simon91}. 
That is, the mesh can be associated to a dual graph, such that the vertices correspond to the finite elements and the edges represent the connectivity of the elements which share common boards. 
If a graph is connected, then it is shown that the second eigenvalue of its Laplacian matrix is positive~\cite{AlgebraicConnectivityGraphs}. 
The components of the second eigenvector are associated with the corresponding vertices of the graph and can be used to assign weights for partitioning of the graph. 

In complex problems with many variables and values, minute errors in obtaining the eigenvalue and eigenvector or a matrix determinant, in calculating an average, a standard deviation or a correlation coefficient, can lead to erroneous decisions. 
Computational platforms offer libraries and functions for carrying out these calculations. 
When it comes to modelling large problems, with complex variables, good, or at least controlled, responses are fundamental.

Little attention has been drawn to assess these platforms under the diversity of operational systems and hardware considering the accuracy of the results.  
Examples of such assessments are found in \cite{AlmironBJPS2009, Almiron2010,Bustosa2006, Keeling2007}. 
Most of these studies, usually limited to spreadsheets, follow the methodology suggested by McCullough \cite{McCullough1998, McCullough1999a,McCullough2002, McCullough2005, McCulluogh2000}: constasting results with the certified values provided by the Statistical Reference Datasets (StRD) produced by the National Institute of Standards and Technology (NIST)~\cite{National2010}.
We add a Monte Carlo study to the protocol in some of the datasets in order to verify the stability of the results with respect to small departures from the original input.
Besides statistical tests, in~\cite{FreryCilamce2010} we proposed additional tests that employ operations on matrices in order to assess the scientific platforms from this viewpoint.

In this work we test three numerical scientific platforms: Octave~3.2.4, Scilab~5.3  and Matlab~R2011a, under the three well known operating systems: Windows~XP Professional SP~2, Linux Ubuntu~10.4 and Mac~OS~X Leopard~10.5.6, whenever the former are avaliable. 
In all cases, i386 architecture hardware was employed, and double precision computation was enforced. 

\medskip
\paragraph{Outline.}
The remainder of this work is organized as follows. 
Section~\ref{sec:MeasuringAccuracy} discusses how accuracy is measured. 
Section~\ref{sec:Results} presents the results obtained assessing basic statistics (subsection~\ref{sec:BasicStatistics}), probability distribution functions (subection~\ref{sec:StatisticalFunctions}), linear regression (subsection~\ref{sec:LinearRegression}) and  operations on matrices (subsection~\ref{sec:Matrices}). 
Section~\ref{sec:Conclusions} concludes the paper.

\section{Measuring Accuracy}\label{sec:MeasuringAccuracy}

Errors in computational simulations can occur and arise from diferents sources. 
They range from modelling errors, defined by the difference between real world and the computable model, to numerical errors introduced in the solution of the problem. 
The latter are (i)~round-off errors, (ii)~truncation and discretization errors, or (iii)~numerical instability. 
Usually the availability to implementation details of the algorithms is very limited and, even when available, other factors, like hardware, compiler, operational system, may compromise the software accuracy.

Considering such limitations, many authors (see, for instance, \cite{AlmironBJPS2009, Almiron2010, Bustosa2006, Keeling2007}) adopt the strategy of measuring the software accuracy from the user viewpoing, that is, comparing the results provided by the software with certified values known to be correct. 
The certified values and datasets employed in this study are obtained in the \emph{Statistical Reference Datasets} from the National Institute of Standards and Technology (NIST)~\cite{National2010}.
We also measure the stability of the results by Monte Carlo, and propose a strategy that considers the results of operations on matrices. 

In the first strategy, statistical descriptive measures are assessed: the  mean, the standard deviation and the first lag coefficient of autocorrelation. 
Linear regression and quantiles of tail probabilities of usual distributions are also computed. 

The LRE (base-10 logarithm of the absolute value of the relative error) is computed as a measure of the accuracy of the functions. 
LRE is approximately the number of matching significant digits between the certified and obtained values: 
\begin{equation}
\label{MD}
\text{LRE}(x,c)=\left\{ \begin{array}{r l}
			   	-\log_{10} \frac{| x-c|}{| c|}  & \quad\mbox{if}\quad c\neq 0,	\\
			  	-\log_{10} | x|   & \quad\mbox{otherwise},
			  \end{array} \right.
\end{equation}
where  $x$ is the result of evaluation function computed by the software under assessment and $c$ is the certified value. 

The following convention was adopted:  when $\text{LRE}(x,c)\geq 1$ we consider only one decimal place.  
If   $0\leq LRE(x,c)< 1$, it is assumed zero, that is, no correct digit was found. 
If the value was very far from the certified, ``--'' is used; the word `Inf' is used to mean that there is a perfect match, and when the platform returns an error, it is denoted by `NA'. 

LREs were computed using the R platform (\url{http://www.r-project.org/}), whose excellent numerical properties were checked in~\cite{AlmironBJPS2009}.

In order to assess the stability of the results, we propose a Monte Carlo procedure for some of the datasets used to compute the precision of statistical descriptive measures.
A bootstap estimate of the $LRE$ produced by each platform for each measure (mean, standard deviation and first lag coefficient of autocorrelation) was computed for each of the four real-world datasets plus \texttt{PiDigits}.
One hundred independent vectors of the same size of the original ones were obtained sampling with reposition from each original data set.
Certified values were computed using R.
These vectors were submitted to each platform, and the resulting observed quantities were contrasted with the certified values producing $100$ $LRE$s for each quantity of interest.
These last values were used to compute an estimate of the standard deviation: 
$$
s_{LRE}=\sqrt{\frac1{99}\sum_{r=1}^{100}[LRE(r)-LRE]^2},
$$
where $LRE$ denotes the ``true'' logarithm of the absolute value of the relative error, which was observed with the original dataset.
This procedure does not belong to the original protocol, but it allows verifying how stable the resuts are with respect to relatively small perturbations of the dataset.
 
The second strategy deals with operations on matrices. 
We propose two assessments: the first is to compute the determinant of a $2\times 2$ matrix whose certified value is zero $|M|=0$. 
Consider the matrix
$$
M = \left(
  \begin{array}{cc}
   b & b\varepsilon\\
  s/\varepsilon & s\\
  \end{array}
\right),
$$
with arbitrary values $b,s,\epsilon$.

The numerical computation of the determinant with built-in functions will guarantee that the intermediate values $(b\varepsilon)$ and $(s/\varepsilon)$ are evaluated. 
The values we proposed to be assessed are $b=10^j$ and $s=10^{-j}$, with $j\in{0,1,\dots,15}$, and 
$$
\varepsilon = 0.\underbrace{9\cdots9}_{k \text{ times}}, \quad k\in\{1,\dots,15\}.
$$ 
The measure of accuracy  is the result of the logical comparison of the computed value and zero in the platform under assessment. 
That is, the interest in such case is not the value itself but the result of comparison.

We also propose another assessment considering spectral graph theory. 
The interest in this proposal is that the Laplacian matrix is directly related to many properties of the graph as, for instance, connectivity~\cite{Bollobas1998}.

Let $G=(V,E)$ be a non-directed finite graph without loops such that $V=\{w_1,w_2,\dots,w_n\}$ is the set of vertices and $E$ is the set of edges. 
Denote by $\operatorname{deg}(w_i)$ the degree of vertex $w_i$. 
Let $D$ be the diagonal degree matrix with entries $\operatorname{deg}(w_i)$ and $A$ be the adjacency matrix with elements  $a_{ij}$, which take value $1$ if there is an edge between $w_i$ and $w_j$. 
The Laplacian matrix $\mathcal L(G)$ of $G$ is the difference between $D$ and $A$, i.e., $\mathcal L(G)=D-A$. Fiedler~\cite{AlgebraicConnectivityGraphs} noted that: 
\begin{itemize}
 \item The number of zero eigenvalues of $\mathcal L(G)$ is the number of connected components in the graph;
 \item If $\lambda_1,\dots, \lambda_n$ are the eigenvalues of $\mathcal L(G)$, then $0=\lambda_1\leq \lambda_2 \leq \cdots \leq \lambda_n$.
  \item If the second smallest eigenvalue $\lambda_2$ is greater then zero, $\lambda_2>0$, then $G$ is connected and $\lambda_2$ is called \emph{algebraic connectivity}. 
\end{itemize}

Considering algebraic connectivity, we propose an accuracy assessment based on the class of complete bipartite graphs. 
In such graphs we have two subsets of vertices, say $V_1$ and $V_2$, such that no connections exists between vertices belonging to the same subset, and each vertex from $V_1$ is connected to every vertex from $V_2$. 
Let $m$ and $n$ be the cardinality of $V_1$ and $V_2$, respectively.  
If we denote this bipartite graph by  $K_{m,n}$, its Laplacian matrix has the following form:
\begin{equation}
\mathcal L(K_{m,n}) = \left(
\begin{array}{ccccccc}
n & 0 & \cdots & 0 & -1 & \cdots & -1\\
0 & n & \cdots & 0 &  -1 & \cdots & -1\\
\vdots & &\ddots  & \vdots & \vdots & & \vdots \\
0 & \cdots & 0 & n &  -1 & \cdots & -1\\
-1 & \cdots & -1 & -1 & m & \cdots & 0 \\ 
\vdots & &\ddots  & \vdots & \vdots & & \vdots \\
-1 & \cdots & -1 & -1 & 0 & \cdots & m \\ 
\end{array}
\right).\label{eq:Laplacians} 
\end{equation}
For this Laplacian matrix, Bolloboas~\cite{Bollobas1998} showed that the eigenvalues are $\lambda_1=0$,  $\lambda_{m+n}=m+n$ and there are $n-1$ eigenvalues whose value is $m$, and  $m-1$ eigenvalues whose value is $n$. 

In order to do the assessment, we considered two special cases: one with almost perfect balance $K_{m,m+1}$, and other with almost the worst possible balance $K_{2,2m-1}$, where $m\in\{9,99,999\}$. 
We formed examples of three sizes of graphs: small, medium and big. 
The assessment is based upon the observation of seven quantities:
\begin{itemize}
\item[(i)] the LRE of the smallest eigenvalue ($\lambda_1=0$) denoted $\ell_1$,
\item[(ii)]the LRE of the biggest eigenvalue ($\lambda_{m+n}=m+n$) denoted $\ell_{m+n}$,
\item[(iii)] the LRE of the sum of the eigenvalues ($\sum_i^{m+n} \lambda_i=2mn$) denoted $\ell_S$,
\item[(iv)] the minimum LRE of those eigenvalues that should take value $n$ (there are $m-1$ of them)  denoted $\ell_n$,
\item[(v)] the minimum LRE of those eigenvalues that should take value $m$ (there are $n-1$ of them) denoted $\ell_m$, and
\item[(vi,vii)] the percentage of eigenvalues which test equal to $m$ and to $n$ (being the correct answers $n-1$ and $m-1$, respectively)  denoted $\ell_N$ and $\ell_M$.
\end{itemize}

\section{Results}\label{sec:Results}

In this section we present the results of applying the two strategies described in the previous section. 
The tables present the accuracy of the three programming ambients under assessment running (whenever available) under Windows (`Win'), Linux (`Lin') and Mac~OS (`Mac'), in 32 and 64 bits architecture.   

\subsection{Basic Statistics}\label{sec:BasicStatistics}

The univariate summary statistics we assessed are the sample mean, the sample standard deviation and the sample first lag correlation of nine datasets.
These datasets are classified by NIST in three levels of numerical difficulty: \emph{low}, \emph{average} and \emph{high}. 
The datasets with low difficulty are \texttt{Lew}, \texttt{Lottery}, \texttt{Mavro}, \texttt{Michelso} (these four datasets come from real world experiments), \texttt{NumAcc1} and \texttt{PiDigits}. 
The average difficulty datasets are \texttt{NumAcc2} and \texttt{NumAcc3}, while \texttt{NumAcc4} is the only high difficulty dataset.
The certified values were calculated using multiple precision arithmetic to obtain $500$ digits answers.

The command  \verb+mean+ is common to all platforms. 
The standard deviation in Octave and MatLab is computed with the command \verb+std+, whereas the Scilab command is \verb+st_deviation+.
For computing the correlation, Scilab provides the function \verb+correl+ which, surprisingly and in spite of what is informed in the documentation, returns the covariance rather than the correlation; the correlation was obtained dividing this result by the product of the sample standard deviations of the subvectors.
In Octave we used the command \verb+Correl(v(1:n-1), v(2:n))+, and  in Matlab the command applied was \verb+corr(v(1:n-1), v(2:n))+, considering in both the vector \verb+v+ of size $n\geq3$.

The values in parenthesis are the bootstrap estimates of the standard deviation of $LRE$s, $s_{LRE}$.
Whenever `Inf' was observed, $LRE=16$, i.e., the highest possible accuracy in double precision, was used.

\begin{table}[hbt]
\tiny
\centering
\setlength{\tabcolsep}{1pt}
\caption{LREs for the basic statistics and bootstrap estimates of selected standard deviations}\label{tab:SampleLRE}
 \begin{tabular}{rrrrrrrrrrr}
\toprule
Platform & OS & \begin{sideways}\texttt{PiDigits}\end{sideways} 
& \begin{sideways}\texttt{Lottery}\end{sideways} 
& \begin{sideways}\texttt{Lew}\end{sideways} 
& \begin{sideways}\texttt{Mavro}\end{sideways} 
& \begin{sideways}\texttt{Michelso}\end{sideways} 
& \begin{sideways}\texttt{Numacc1}\end{sideways} 
& \begin{sideways}\texttt{Numacc2}\end{sideways} 
& \begin{sideways}\texttt{Numacc3}\end{sideways} 
& \begin{sideways}\texttt{Numacc4}\end{sideways} \\ \midrule
& & \multicolumn{9}{c}{Sample Mean}\\ \midrule
\multirow{3}{*}{Octave} & Windows & Inf(0) & Inf(0.99) & Inf(0) & Inf(0.04) & Inf(0.04) & Inf & Inf & Inf & Inf\\
&Linux &Inf(0.04) & Inf(10.44) & Inf(11.45) & Inf(11.35) & Inf(10.94) & Inf & Inf & Inf & Inf \\
& Mac Os & Inf(0) & Inf(0.99) & Inf(0) & Inf(0.33) & Inf(0.43) & Inf & Inf & 14.0 & Inf \\ \midrule
\multirow{2}{*}{MatLab} & Windows & 16.0(0) & 15.1(0.89) & 16.0(0) & 16.0(0.35) & 15.4(10.87) & 16.0 & 14.0 & 15.0 & 13.9 \\
&Linux & 16.0(0) & 16.0(0.03) & 16.0(0) & 16.0(0.35) & 16.0(11.48) & 16.0 & 14.0 & 14.0 & 13.9 \\ \midrule
\multirow{3}{*}{Scilab} & Windows & Inf(0.04) & 8.1(6.28) & Inf(0) & Inf(0.04) & Inf(0.04) & Inf & Inf & Inf & 7.7 \\
& Linux & Inf(0) & 8.1(7.92) & Inf(0) & Inf(0.33) & Inf(0.43) & Inf & Inf & Inf & 7.7 \\
& Mac OS & Inf(0) & 8.1(7.92) & Inf(0) & Inf(0.33) & Inf(0.43) & Inf & Inf & Inf & 7.7 \\ \midrule
& & \multicolumn{9}{c}{Sample Standard Deviation}\\ \midrule
\multirow{3}{*}{Octave} & Windows & Inf(0.88) & Inf(0.24) & Inf(0.87) & 13.1(2.94) & 13.9(1.96) & Inf & Inf & 9.5 & 8.3 \\
&Linux & Inf(11.08) & Inf(9.88) & Inf(10.80) & 13.1(10.88) & 13.9(10.88) & Inf & Inf & 8.3 & Inf \\
& Mac Os & Inf(1.76) & Inf(0.35) & Inf(0.42) & 13.1(2.75) & 13.8(1.99) & Inf & Inf & 9.5 & 8.3 \\ \midrule
\multirow{2}{*}{MatLab} & Windows & 14.8(0.70) & 16.0(0.35) & 15.2(0.53) & 13.8(2.06) & 13.9(12.85) & 16.0 & 16.0 & 9.4 & 8.2 \\
&Linux & 14.8((0.70) & 16.0(0.35) & 16.0(0.41) & 13.8(2.06) & 13.9(12.85) & 16.0 & 16.0 & 9.4 & 8.2 \\ \midrule
\multirow{3}{*}{Scilab} & Windows & 7.9(4.79) & 8.1(6.12) & 8.2(6.01) & 4.1(6.33) & 6.2(6.5) & Inf & Inf & Inf & Inf  \\
& Linux & 7.9(8.01) & 8.1(7.67) & 8.2(7.5) & 4.1(11.79) & 6.2(9.62) & Inf & Inf & Inf & Inf  \\
& Mac OS & 7.9(6.47) & 8.1(7.67) & 8.2(7.41) & 4.1(11.59) & 6.2(9.62) & Inf & Inf & Inf & Inf \\ \midrule
& & \multicolumn{9}{c}{Sample First Lag Correlation}\\ \midrule
\multirow{3}{*}{Octave} & Windows & 4.0(0.48) & 2.1(0.67) & 2.6(0.75) & 1.8(0.56) & 3.6(1.90) & 0 & 3.0 & 3.0 & 3.0 \\
&Linux & 4.0(2.10) &2.1(0.60) & 2.6(0.50) & 1.8(0.55) & 3.6(1.81) & 0 & 3.3 & 3.3 & 3.3  \\
& Mac Os & 4.0(1.05) & 2.1(0.67) & 2.6(0.75) & 1.8(0.58) & 3.6(1.52) & 0 & 3.0 & 3.0 & 3.0 \\ \midrule
\multirow{2}{*}{MatLab} & Windows & 3.9(6.06) & 2.0(3.49) & 2.6(4.08) & 1.7(2.78) & 3.5(4.82) & 0 & 3.3 & 3.3 & 3.3 \\ 
&Linux & 3.9(6.06) & 2.0(3.49) & 2.6(4.08) & 1.7(2.78) & 3.5(4.82) & 0 & 3.3 & 3.3 & 3.3 \\ \midrule
\multirow{3}{*}{Scilab} & Windows & -- & -- & -- & 0(1.75) & 0(2.09) & 0.5 & 0 & 0 & 0  \\
& Linux & -- & -- & -- & 0(2.09) & 0(2.09) & 0.5 & 0 & 0 & 0  \\
& Mac OS & -- &  0(2.39) & 0(2.40) & 0(1.75) & 0(2.09) & 0.5 & 0 & 0 & 0  \\ \bottomrule
\end{tabular}
\end{table}

\subsection{Statistical Functions}\label{sec:StatisticalFunctions}

The distributions herein assessed are the binomial (Table~\ref{tab:binomial}), Poisson (Tables~\ref{tab:Poisson1} and~\ref{tab:Poisson2}), gamma (Table~\ref{tab:Gamma}), normal (Table~\ref{tab:normal}), $\chi^2$ (Table~\ref{tab:chisquare}), beta (Table~\ref{tab:beta}), t-Student (Table~\ref{tab:t}) and $F$ (Table~\ref{tab:F}).  

The commands to compute all the distributions, except for the beta in Matlab and Octave, are the same: \verb+binocdf+ for the binomial,  \verb+poisscdf+ for the Poisson, \verb+gamcdf+ for the gamma, \verb+norminv+ for the normal, \verb+cdfchi+ for the $\chi^2$, \verb+tinv+ for the t-Student, and \verb+finv+ for the $F$ law. 
The command \verb+beta_cdf+ computes beta distribuition in Octave, while Matlab provides the command \verb+betainv+. 
Scilab provides \verb+cdfbin+, \verb+cdfpoi+, \verb+cdfgam+, \verb+cdfnor+, \verb+cdfchi+, \verb+cdfbet+, \verb+cdft+ and \verb+cdff+ for computing the binomial, Poisson, gamma, normal, $\chi^2$,  beta, t-Student and $F$ quantiles, respectively. 

\begin{table}[hbt]
\tiny
\centering
\caption{Binomial distribution, $n = 1030$ and $p = 1/2$}\label{tab:binomial}	
\begin{tabular}{rrrrrrrrrr}
\toprule
 & \multicolumn{1}{c}{$\Pr(X \leq k)$} & \multicolumn{2}{c}{Matlab} & \multicolumn{3}{c}{Octave} & \multicolumn{3}{c}{Scilab}\\ 
\cmidrule(lr{.25em}){3-4} \cmidrule(lr{.25em}){5-7} \cmidrule(lr{.25em}){8-10}
$k$ & \multicolumn{1}{c}{Certified} & \multicolumn{1}{c}{Win} & \multicolumn{1}{c}{Lin} & \multicolumn{1}{c}{Win} & \multicolumn{1}{c}{Lin} & \multicolumn{1}{c}{Mac} & \multicolumn{1}{c}{Win} & \multicolumn{1}{c}{Lin} & \multicolumn{1}{c}{Mac}\\ \midrule
1 & 8.96114E-308 & 3.0 & 3.0 & 0 & 0 & 0 & Inf & Inf & Inf \\
2 & 4.61499E-305 & 3.0 & 3.0 & 0 & 0 & 0 & 8.0 & 8.0 & 8.0 \\
100 & 1.39413E-169 & 1.0 & 1.0 & 0 & 0 & 0 & 7.0 & 7.0 & 7.0 \\
300 & 2.91621E-42 & 0 & 0& 0 & 0 & 0 & 7.0 & 7.0 & 7.0 \\
400 & 3.89735E-13 & 0 &0 & 0 & 4.0 & 4.0 & 6.0 & 6.0 & 6.0 \\
410 & 3.19438E-11 & 0 &0 & 0 & 6.0 & 6.0 & 6.0 & 6.0 & 6.0 \\
\bottomrule
 \end{tabular}
\end{table}

\begin{table}[hbt]
\tiny
\centering
\caption{Poisson probabilities, $\lambda=200$}\label{tab:Poisson1}	
 \begin{tabular}{rrrrrrrrrr}
 \toprule
 & \multicolumn{1}{c}{$\Pr(X = k)$} & \multicolumn{2}{c}{Matlab} & \multicolumn{3}{c}{Octave} & \multicolumn{3}{c}{Scilab}\\ 
\cmidrule(lr{.25em}){3-4} \cmidrule(lr{.25em}){5-7} \cmidrule(lr{.25em}){8-10}
$k$ & \multicolumn{1}{c}{Certified} & \multicolumn{1}{c}{Win} & \multicolumn{1}{c}{Lin} & \multicolumn{1}{c}{Win} & \multicolumn{1}{c}{Lin} & \multicolumn{1}{c}{Mac} & \multicolumn{1}{c}{Win} & \multicolumn{1}{c}{Lin} & \multicolumn{1}{c}{Mac}\\ \midrule
0 & 1.38390E-87 & 5.6 & 5.6 & 6.0 & 6.0 & 6.0 & 7.0 & 7.0 & 7.0\\
103 &1.41720E-14 & 1.4 & 1.4 & 1.0 & 1.0 & 1.0 & 2.0 & 0 & 0 \\
315 & 1.41948E-14 & 0 & 0 & 0 & 0 & 0 & 0 & 0 & 0 \\
400 & 5.58069E-36 & 6.4 &  6.4 & 6.0 & 6.0 & 6.0 & 0 & 0 & 0 \\ 
900 & 1.73230E-286 & 6.0 & 6.0 & 6.0 & 6.0 & 6.0 & 0 & 0 & 0 \\
\bottomrule
 \end{tabular}
\end{table}

\begin{table}[hbt]
\tiny
\centering
\caption{Poisson distribution functions, $\lambda=200$}\label{tab:Poisson2}	
 \begin{tabular}{rrrrrrrrrrr}
 \toprule
& & \multicolumn{1}{c}{$\Pr(X \leq k)$} & \multicolumn{2}{c}{Matlab} & \multicolumn{3}{c}{Octave} & \multicolumn{3}{c}{Scilab}\\
 \cmidrule(lr{.25em}){4-5} \cmidrule(lr{.25em}){6-8} \cmidrule(lr{.25em}){9-11}
 $k$ & $\lambda$ & \multicolumn{1}{c}{Certified} & \multicolumn{1}{c}{Win} & \multicolumn{1}{c}{Lin} & \multicolumn{1}{c}{Win} & \multicolumn{1}{c}{Lin} & \multicolumn{1}{c}{Mac} & \multicolumn{1}{c}{Win} & \multicolumn{1}{c}{Lin} & \multicolumn{1}{c}{Mac}\\ \midrule
1E+05 & 1E+05 & 0.500841 & 1.0 & 1.0 & 1.0 & 1.2 & 1.0 & 7.0 & 7.0 & 7.0 \\
1E+07 & 1E+07 & 0.500084 & 7.0 & 7.0 & 7.0 & 7.0 & 7.0 & 6.0 & 6.0 & 6.0 \\
1E+09 & 1E+09 & 0.500008 & 7.0 & 7.0 &  7.0 & 7.0 & 7.0 & 6.0 & 6.0 & 6.0 \\
\bottomrule
 \end{tabular}
\end{table}

\begin{table}[hbt]
\tiny
\centering
\caption{Gamma distribution functions, $\beta=1$}\label{tab:Gamma}	
 \begin{tabular}{*{11}r}\toprule
& & \multicolumn{1}{c}{$\Pr(X \leq x)$} & \multicolumn{2}{c}{Matlab} & \multicolumn{3}{c}{Octave} & \multicolumn{3}{c}{Scilab}\\ \cmidrule(lr{.25em}){4-5} \cmidrule(lr{.25em}){6-8} \cmidrule(lr{.25em}){9-11}
 $x$ & $\alpha$ & \multicolumn{1}{c}{Certified} & \multicolumn{1}{c}{Win} & \multicolumn{1}{c}{Lin} & \multicolumn{1}{c}{Win} & \multicolumn{1}{c}{Lin} & \multicolumn{1}{c}{Mac} & \multicolumn{1}{c}{Win} & \multicolumn{1}{c}{Lin} & \multicolumn{1}{c}{Mac}\\ \midrule
0.1 & 0.1 & 0.827552 & 7.0 & 7.0 & 7.0 & 7.0 & 7.0 & 7.0 & 7.0 & 7.0 \\
0.2 & 0.1 & 0.879420 & 6.0 & 6.0 & 6.0 & 6.0 & 6.0 & 6.0 & 6.0 & 6.0 \\
0.2 & 0.2 & 0.764435 & 6.0 & 6.0 & 6.0 & 6.0 & 6.0 & 6.0 & 6.0 & 6.0 \\
0.4 & 0.3 & 0.776381 & 6.0 & 6.0 & 6.0 & 6.0 & 6.0 & 6.0 & 6.0 & 6.0 \\
0.5 & 0.4 & 0.748019 & 6.0 & 6.0 & 6.0 & 6.0 & 6.0 & 6.0 & 6.0 & 6.0 \\
\bottomrule
 \end{tabular}
\end{table}

\begin{table}[hbt]
\tiny
\centering
\caption{Normal quantiles, $\mu = 0$ and $\sigma=1$}\label{tab:normal}	
 \begin{tabular}{*{10}r}
 \toprule
 & & \multicolumn{2}{c}{Matlab} & \multicolumn{3}{c}{Octave} & \multicolumn{3}{c}{Scilab}\\ \cmidrule(lr{.25em}){3-4} \cmidrule(lr{.25em}){5-7} \cmidrule(lr{.25em}){8-10}
$p$ & \multicolumn{1}{c}{Certified $z_p$} & \multicolumn{1}{c}{Win} & \multicolumn{1}{c}{Lin} & \multicolumn{1}{c}{Win} & \multicolumn{1}{c}{Lin} & \multicolumn{1}{c}{Mac} & \multicolumn{1}{c}{Win} & \multicolumn{1}{c}{Lin} & \multicolumn{1}{c}{Mac}\\ \midrule
5E-1 & 0 & Inf & Inf & Inf & Inf & Inf & 0 & 0 & 0 \\
1E-198 & -30.0529 & 7.0 & 7.0 & 7.0 & 7.0 & 7.0 & 3.0 & 3.0 & 3.0 \\
1E-300 & -37.0471 & 7.0 & 7.0 & 7.0 & 7.0 & 7.0 & 1.0 & 1.0 & 1.0 \\
\bottomrule
 \end{tabular}
\end{table}

\begin{table}[hbt]
\tiny
\centering
\caption{The $\chi^2$ distribution}\label{tab:chisquare}	
 \begin{tabular}{*{11}r}
 \toprule
& & \multicolumn{1}{c}{$\Pr(X > x)=p$} & \multicolumn{2}{c}{Matlab} & \multicolumn{3}{c}{Octave} & \multicolumn{3}{c}{Scilab}\\ 
\cmidrule(lr{.25em}){4-5} \cmidrule(lr{.25em}){6-8} \cmidrule(lr{.25em}){9-11}
$p$ & $n$ & Certified $x$ & \multicolumn{1}{c}{Win} & \multicolumn{1}{c}{Lin} & \multicolumn{1}{c}{Win} & \multicolumn{1}{c}{Lin} & \multicolumn{1}{c}{Mac} & \multicolumn{1}{c}{Win} & \multicolumn{1}{c}{Lin} & \multicolumn{1}{c}{Mac}\\ \midrule
2E-1 & 1 & 1.64237 & 5.6 & 5.6 & 5.6 & 5.6 & 5.6 & 5.3 & 5.3 & 5.3 \\
1E-7 & 1 & 28.3740 & 6.4 & 6.4 & 6.4 & 6.4 & 6.4 & 6.4 & 6.4 & 6.4 \\
1E-7 & 5 & 40.8630 & 6.3 & 6.3 & 6.3 & 6.3 & 6.3 & 6.3 & 6.3 & 6.3 \\
1E-12 & 1 & 50.8441 & 6.6 & 6.6 & 7.1 & 7.1 & 7.1 & 6.3 & 6.3 & 6.3 \\
0.48 & 778 & 779.312 & 6.3 & 6.3 & 4.4 & 4.4 & 4.4 & 6.2 & 6.2 & 6.2 \\
0.52 & 782 & 779.353 & 6.3 & 6.3 & 6.3 & 6.3 & 6.3 & 6.3 & 6.3 & 6.3 \\
\bottomrule
 \end{tabular}
\end{table}

\begin{table}[hbt]
\tiny
\centering
\caption{Beta quantiles, $\alpha = 5$ and $\beta=2$}\label{tab:beta}	
 \begin{tabular}{*{10}r}\toprule
 & & \multicolumn{2}{c}{Matlab} & \multicolumn{3}{c}{Octave} & \multicolumn{3}{c}{Scilab}\\ \cmidrule(lr{.25em}){3-3} \cmidrule(lr{.25em}){4-6} \cmidrule(lr{.25em}){7-9}
$p$ & \multicolumn{1}{c}{Certified} & \multicolumn{1}{c}{Win} & \multicolumn{1}{c}{Lin} & \multicolumn{1}{c}{Win} & \multicolumn{1}{c}{Lin} & \multicolumn{1}{c}{Mac} & \multicolumn{1}{c}{Win} & \multicolumn{1}{c}{Lin} & \multicolumn{1}{c}{Mac}\\ \midrule
1E-2 & 2.94314E-01 & 6.0 & 6.0 & 6.0 & 6.0 & 6.0 & 6.0 & 6.0 & 6.0 \\
1E-3 & 1.81386E-01 & 6.1 & 6.1 & 6.0 & 6.0 & 6.0 & 6.0 & 6.0 & 6.0 \\
1E-4 & 1.12969E-01 & 5.4 & 5.4 & 5.0 & 5.0 & 5.0 & 5.0 & 5.0 & 5.0 \\ 
1E-5 & 7.07371E-02 & 6.2 & 6.2 & 6.0 & 6.0 & 6.0 & 6.0 & 6.0 & 6.0 \\ 
1E-6 & 4.44270E-02 & 6.0 & 6.0 & 6.0 & 6.0 & 6.0 & 6.0 & 6.0 & 6.0 \\ 
1E-7 & 2.79523E-02 & 5.9 & 5.9 & 6.0 & 6.0 & 6.0 & 6.0 & 6.0 & 6.0 \\ 
1E-8 & 1.76057E-02 & 6.3 & 6.3 & 6.0 & 6.0 & 6.0 & 6.0 & 6.0 & 6.0 \\
1E-9 & 1.10963E-02 & 5.5 & 5.5 & 5.0 & 5.0 & 5.0 & 5.0 & 5.0 & 5.0 \\
1E-10 & 6.99645E-03 & 6.7 & 6.7 & 7.0& 7.0 & 7.0 & 7.0 & 7.0 & 7.0 \\
1E-11 & 4.41255E-03 & 6.7 & 6.7 & 7.0& 7.0 & 7.0 & 7.0 & 7.0 & 7.0 \\
1E-12 & 2.78337E-03 & 5.9 & 5.9 & 6.0 & 6.0 & 6.0 & 6.0 & 6.0 & 6.0 \\
1E-13 & 1.75589E-03 & 6.1 & 6.1 & 6.0 & 6.0 & 6.0 & 6.0 & 6.0 & 6.0 \\
1E-100 & 6.98827E-21 & 6.8 & 6.8 & 7.0& 0 & 0 & 7.0& 7.0 & 7.0 \\
\bottomrule
 \end{tabular}
\end{table}

\begin{table}[hbt]
\tiny
\centering
\caption{The t-Student distribution, $n=1$}\label{tab:t}	
 \begin{tabular}{*{10}r}\toprule
 & $\Pr(X>x)=p$ & \multicolumn{2}{c}{Matlab} & \multicolumn{3}{c}{Octave} & \multicolumn{3}{c}{Scilab}\\ \cmidrule(lr{.25em}){3-4} \cmidrule(lr{.25em}){5-7} \cmidrule(lr{.25em}){8-10}
$p$ & \multicolumn{1}{c}{Certified $x$} & \multicolumn{1}{c}{Win} & \multicolumn{1}{c}{Lin} & \multicolumn{1}{c}{Win} & \multicolumn{1}{c}{Lin} & \multicolumn{1}{c}{Mac} & \multicolumn{1}{c}{Win} & \multicolumn{1}{c}{Lin} & \multicolumn{1}{c}{Mac}\\ \midrule
1E-8 & 3.18310E+07 & 0 & 0 & 0 & 0 & 0 & 6.0 & 6.0 & 6.0 \\
1E-11 & 3.18310E+10 & 0 & 0 & 0 & 0 & 0 & 6.0 & 6.0 & 6.0 \\ 
1E-12 & 3.18310E+11 & -- & -- &  -- & -- & -- & 6.0 & 6.0 & 6.0 \\
1E-13 & 3.18310E+12 & 0 & 0 & 0 & 0 & 0 & 6.0 & 6.0 & 6.0 \\
1E-100 & 3.18310E+99 & -- & -- &  -- & -- & -- & 8.0 & 8.0 & 0 \\
\bottomrule
 \end{tabular}
\end{table}

\begin{table}[hbt]
\tiny
\centering
\caption{The $F$ distribution, $n_1=n_2=1$}\label{tab:F}	
 \begin{tabular}{*{10}r}\toprule
 & $\Pr(X>x)=p$ & \multicolumn{2}{c}{Matlab} & \multicolumn{3}{c}{Octave} & \multicolumn{3}{c}{Scilab}\\ \cmidrule(lr{.25em}){3-4} \cmidrule(lr{.25em}){5-7} \cmidrule(lr{.25em}){8-10}
$p$ & \multicolumn{1}{c}{Certified $x$} & \multicolumn{1}{c}{Win} & \multicolumn{1}{c}{Lin} & \multicolumn{1}{c}{Win} & \multicolumn{1}{c}{Lin} & \multicolumn{1}{c}{Mac} & \multicolumn{1}{c}{Win} & \multicolumn{1}{c}{Lin} & \multicolumn{1}{c}{Mac}\\ \midrule
1E-5 & 4.05285E+09 & 6.0 & 6.0 & 6.0 & 6.0 & 6.0 & 6.0 & 6.0 & 6.0 \\
1E-6 & 4.05285E+11 & 6.0 & 6.0 & 6.0 & 6.0 & 6.0& 6.0 & 6.0 & 6.0 \\
1E-12 & 4.05285E+23 & 4.0 & 4.0 &4.0 & 4.0 &4.0 & 6.0 & 6.0 & 6.0 \\
1E-13 & 4.05285E+25 & 3.0 & 3.0 & 3.0 & 3.0 & 3.0 & 3.0 & 3.0 & 3.0 \\
1E-100 & 4.05285E+199 & -- & -- & Inf & Inf & Inf & 0 & 0 & 0 \\
\bottomrule
 \end{tabular}
\end{table}

\subsection{Linear Regression}\label{sec:LinearRegression}

NIST offers eleven datasets to perform linear regression analysis. 
The datasets are divided into numerical difficulty levels: two of low level, (\texttt{Norris} and \texttt{Pontius}), two of average level,  (\texttt{Noint1} and \texttt{Noint2}) and seven of high level.
Table~\ref{tab:LinearRegression} presents the smallest LRE of each regression, and the LRE of the residual standard deviation (RSD) of each fit.

Octave and Matlab do not provide explicit functions for performing linear regression.
Rather than that, linear regression is computed solving a least squares problem, and the data requires prior preparation for that. 
Scilab provides the function \verb+reglin+ to obtain the coefficients and RSD.

\begin{table}[hbt]
\tiny
\vskip 35em 
\setlength{\tabcolsep}{1.7pt}
 \caption{LRE of linear regression results}\label{tab:LinearRegression}
\centering
\begin{tabular}{*{17}r}\toprule
 & \multicolumn{4}{c}{Matlab} & \multicolumn{6}{c}{Octave} & \multicolumn{6}{c}{Scilab}\\
\cmidrule(lr{.25em}){2-5} \cmidrule(lr{.25em}){6-11} \cmidrule(lr{.25em}){12-17} %
Data & \multicolumn{2}{c}{Windows} &  \multicolumn{2}{c}{Linux} & \multicolumn{2}{c}{Windows} & \multicolumn{2}{c}{Linux} & \multicolumn{2}{c}{Mac OS} & \multicolumn{2}{c}{Windows} & \multicolumn{2}{c}{Linux} & \multicolumn{2}{c}{Mac OS}\\ 
\cmidrule(lr{.25em}){2-3} \cmidrule(lr{.25em}){4-5} \cmidrule(lr{.25em}){6-7} \cmidrule(lr{.25em}){8-9} \cmidrule(lr{.25em}){10-11} \cmidrule(lr{.25em}){12-13} \cmidrule(lr{.25em}){14-15} \cmidrule(lr{.25em}){16-17}  
  & $\widehat{\beta}$ & RSD & $\widehat{\beta}$ & RSD & $\widehat{\beta}$ & RSD & $\widehat{\beta}$ & RSD & $\widehat{\beta}$ & RSD & $\widehat{\beta}$ & RSD & $\widehat{\beta}$ & RSD & $\widehat{\beta}$ & RSD \\ \midrule
\texttt{Filip} 		& 7.1 & 8.2 & 7.1 & 8.2	& 0 & 1.1 & 0 & 1.1 & 0 & 1.1 		& 0  & 0 & 0 & 0 & 0 &  0\\
\texttt{Longley}  	& 0 & 0 & 0 & 0	& 0 & 0 & 0 & 0 & 0 &  0		& 0 & 0 & 0 & 0 & 0 & 0 \\
\texttt{Norris}  	& 13.4 & 14.1 & 13.4 & 14.1	& 0 & 1.8 & 12.1 & 1.8 & 12.3 &  1.8	& 7.9  & 1.3 & 8 & 1.3 & 10.0 & 1.6 \\
\texttt{Noint1}  	& 0 & 0	 & 0 & 0	& Inf & 2.8 & Inf & 2.8 & Inf & 2.9 	& 0 & 0 & 0 & 0 & 0 & 0 \\
\texttt{Noint2}  	& 0 & 0	 & 0 & 0	& Inf & 2.3 & Inf & 2.3 & Inf &  2.3	& 0 & 0 & 0 & 0 & 0 & 0 \\
\texttt{Pontius}  	& 3.6 & 13.2 & 3.6 & 13.2   & 6.4 & 1.6 & 7.1 & 1.6 & 6.2 & 1.6  & 8.0 & 1.4 & 8.0 & 1.4 & 5.5 & 1.4 \\
\texttt{Wampler1}  	& 9.3 & 9.4 & 9.3 & 9.4   & 7.0 & 4.8 & 6.9 & 4.6 & 6.6 & 4.4 	& 4.0 & 3.5 & 3.1 & 3.5 & 2.7 & 4.1 \\
\texttt{Wampler2}  	& Inf & 14.2	& Inf & 14.2	& 9.6 & 9.3 & 9.8 & 9.5 & 9.8 &  0 &  6.9 & 8.5 & 6.2 & 8.5 & 6.0 & 8.3 \\
\texttt{Wampler3}  	& 0 & 14.2 & 0 & 14.2 & 6.9 & 0 & 7.0 & 0 & 6.6 & 0 & 2.9 & 0 & 3.5 & 0 & 2.7 & 0 \\
\texttt{Wampler4}  	& 0 & 1.6	& 0 & 1.6	& 6.9 & 0 & 7.0 & 0 & 6.6 &  0 & 2.9 & 0 & 3.5 & 0 & 2.7  & 0 \\
\texttt{Wampler5}  	& 0 & 1.6	& 0 & 1.6	& 6.5 & 0 & 7.0 & 0 & 6.2 & 0 & 2.9 & 0 & 3.5 & 0 & 2.7 & 0 \\
\bottomrule
\end{tabular}
\end{table}

\subsection{Results on Decisions based on Matrices}\label{sec:Matrices}

The command \verb+det+ is used by all three platforms under assessment to compute determinants. 
As proposed Section~\ref{sec:MeasuringAccuracy}, the evaluation is based on comparing the results with the certified value zero, rather than on the numerical value itself. 
This is due to the fact that more often than not what users are interested upon is a decision, and not a numerical value.

Curiosily, the number of correct results of comparing $\widetilde{|M|}$ with zero was the same, that is, exactly $146$ for the three platforms under assessment.

The results of computing spectral graph analyses are presented in Table~\ref{tab:GraphOctaveSciLab}.

\begin{sidewaystable}
\centering
\tiny
\setlength{\tabcolsep}{1.7pt}
\vskip 45em 
 \caption{Accuracy computing spectral graph analyses}\label{tab:GraphOctaveSciLab}
\begin{tabular}{*{22}r}\toprule
& \multicolumn{7}{c}{Windows} & \multicolumn{7}{c}{Linux}  \\
\cmidrule(lr{.25em}){2-8} \cmidrule(lr{.25em}){9-15}   
$m,n$  &$\ell_1$  & $\ell_{m+n}$ & $\ell_S$ & $\ell_n$ & $\ell_m$ & $\ell_N$ & $\ell_M$    &$\ell_1$  & $\ell_{m+n}$ & $\ell_S$ & $\ell_n$ & $\ell_m$ & $\ell_N$ & $\ell_M$  &&&&&&&\\ \midrule
$9,10$ 		& 15.2 & Inf & 15.8 & 14.9 & 14.9 & 11.1 & 62.5 & 15.2 & Inf & 15.8 & 14.9 & 14.9 & 11.1 & 62.5 &&&&&&& \\
$99,100$ 		& 12.7 & 14.8 & Inf & 14.6 & 13.8 & 7.1 & 4.1 & 12.7 & 14.8 & Inf & 14.6 & 13.8 & 7.1 & 4.1 &&&&&&&\\
$999,1000$ 	& 10.5 & 14.5 & 15.1 & 13.8 & 12.6 & 1.3 & 2.3 & 10.5 & 14.5 & 15.1 & 13.8 & 12.6 & 1.3 & 2.3 &&&&&&&\\   \midrule
$2,17$		& 15.3 & Inf & 15.7 & Inf & 14.7 & 12.1 & 100.0  & 15.3 & Inf & 15.7 & Inf & 14.7 & 12.1 & 100.0 &&&&&&&\\
$2,197$		& 14.3 & 15.8 & 15.2 & inf & 11.7 & 6.1 & 100.0  & 14.3 & 15.8 & 15.2 & inf & 11.7 & 6.1 & 100.0 &&&&&&&\\
$2,1997$		& 13.1 & 15.9 & 14.2 & 15.9 & 9.6 & 1.1 & 0  & 13.1 & 15.9 & 14.2 & 15.9 & 9.6 & 1.1 & 0 &&&&&&&\\  \midrule
 & \multicolumn{7}{c}{Windows} & \multicolumn{7}{c}{Linux} &\multicolumn{7}{c}{Mac OS} \\
\cmidrule(lr{.25em}){2-8} \cmidrule(lr{.25em}){9-15} \cmidrule(lr{.25em}){16-22}  
$m,n$  &$\ell_1$  & $\ell_{m+n}$ & $\ell_S$ & $\ell_n$ & $\ell_m$ & $\ell_N$ & $\ell_M$    &$\ell_1$  & $\ell_{m+n}$ & $\ell_S$ & $\ell_n$ & $\ell_m$ & $\ell_N$ & $\ell_M$      &$\ell_1$  & $\ell_{m+n}$ & $\ell_S$ & $\ell_n$ & $\ell_m$ & $\ell_N$ & $\ell_M$    \\
\midrule
\multicolumn{22}{c}{Octave}\\
\midrule
$9,10$ & 15.4 & 15.7 & 15.5 & 15.0 & 14.9 & 37.5 & 11.1 &  15.4 & 15.7 & 15.5 & 15.0 & 14.9 & 11.1 & 37.5	  & 15.2 & Inf & Inf & 14.8 & 15.1 & 12.5 & 22.2 \\
$99,100$ 	& 13.1 & 15.2 & 15.7 & 14.3 & 14.5 & 5.1 & 13.1 	& 13.5 & 15.2 & 15.7 & 14.3 & 14.5 & 5.0 & 9.2	  & 13.0 & 15.3 & 15.4 & 13.9 & 14.1 & 8.0 & 7.1 \\
$999,1000$ & 11.1 & 14.3 & 14.9 & 13.1 & 13.3 & 0.8 &	1.0 	& 11.1 & 14.3 & 14.9 & 13.1 & 13.3 & 0.8 &	1.0	& 11.1 & 14.2 & 14.9 & 12.7 & 12.7 & 1.6 & 0.5\\ \midrule
$2,17$	& 14.9 & 15.4 & Inf & 14.5 & 15.2 & 0 & 6.2   & 14.9 & 15.4 & Inf & 14.5 & 15.2 & 0 & 6.2		& 15.4 & Inf & 15.6 & 14.3 & 15.3 & 0 & 37.5 \\
$2,197$ 	& 14.4 & 15.9 & 14.7 & 12.1 & 14.4 & 0 & 17.8	& 14.4 & 15.9 & 14.7 & 12.1 & 14.4 & 0 & 17.8		& 14.9 & Inf & 14.8 & 12.2 & 14.5 & 0 & 7.6 \\
$2,1997$ 	& 14.1 & 15.6 & Inf & 15.0 & 9.7 & 0 & 4.3   	& 14.1 & 15.6 & Inf & 15.0 & 9.7 & 0 & 4.3		& 13.1 & 15.6 & 14.8 & 9.6 & 14.0 & 0 & 1.9 \\ \midrule
\multicolumn{22}{c}{Scilab}\\
\midrule
$9,10$ & 14.6 & Inf & 15.8 & 15.1 & 14.9 & 37.5 & 44.4 &  15.2 & Inf & 15.8 & 15.4 & 15.0 & 25.0 & 44.0	  & 15.2 & Inf & Inf & 14.8 & 15.1 & 12.5 & 22.2 \\
$99,100$  	& 13.4 & 15.5 & 15.7 & 14.6 & 14.2 & 5.1 & 10.1		& 14.3 & 15.5 & 14.8 & 14.7 & Inf & 7.1 & 6.1	       & 13.0 & 15.4 & 15.7 & 13.8 & 13.8 & 4.1 & 5.0 \\
$999,1000$  & 12.1 & 14.9 & 15.6 & 13.6 & 13.3 & 1.1 & 3.9		& 11.4 & 14.9 & 15.6 & 13.6 & 14.1 & 1.0 & 1.8	   & 10.9 & 14.2 & 14.8 & 13.1 & 13.1 & 1.3 & 1.1 \\ \midrule
$2,17$  	& 15.0 & 15.4 & Inf & 15.5 & 15.2 & 0 &	16.2		& 15.0 & 15.4 & Inf & 15.5 & 15.2 & 0 &	16.2	        & 15.5 & Inf & 15.7 & 15.4 & 14.4 & 0 & 37.5 \\
$2,197$  	& 14.5 & 15.8 & 14.8 & 14.7 & 12.3 & 0 & 20.9		& 14.5 & 15.8 & 14.8 & 14.7 & 12.3 & 0 & 20.9	      & 14.2 & 15.6 & 14.3 & 14.0 & 12.0 & 0 & 17.3 \\
$2,1997$  	& 13.0 & 15.6 & 14.6 & Inf & 9.9 & 100 & 13.5		& 13.0 & 15.6 & 14.6 & Inf & 9.9 & 100 & 13.5 	     & 13.1 & 15.6 & 13.5 & 12.9 & 10.0 & 0 & 8.0 \\ \bottomrule
\end{tabular}
\end{sidewaystable}

\section{Conclusions}\label{sec:Conclusions}

Regarding the computation of basic statistics, Table~\ref{tab:SampleLRE} shows that the mean poses little difficulty for the platforms, with the exception of Octave for Linux, which presented the smallest number of LRE in five of the nine datasets ($\text{LRE}(x,c)\leq 7$).
Surprisingly, these five datasets offer low numerical difficulty.

When computing the standard deviation, Octave presented the bests results when comparing with others two platforms. 
The version tested here was better than the one tested before (see \cite{FreryCilamce2010}). 
Scilab presented an unacceptable low accuracy in a single dataset, for which $\text{LRE}(x,c)\leq 5$.

As in other studies, c.f.\ reference~\cite{AlmironetalJSS2010}, the first-lag sample autocorrelation is a challenging quantity to compute.
None of the platforms here tested provided acceptable results. 
All of them computed $\text{LRE}(x,c)\leq 5$, and Scilab had the worst performance.

Scilab is also the worst platform with respect to the stability of the results, as measured by estimated standard variation of the observed $LRE$.
As can be noted in Table~\ref{tab:SampleLRE}, all conclusions about the standard deviation may be reverted with small perturbations of the original input, e.g., the best results which were produced for the \texttt{Lew} dataset can be turned into unacceptable by subtracting $s_{LRE}$ from the observed $LRE$.

Two other cases are notorious for their instability: the sample mean and the sample standard deviation, both computed by Octave under Linux.
In most of the other cases, small perturbations of the original input do not change the conclusion about the precision.

Scilab presented the best performance when dealing with the binomial and t-Student distributions, and also when computing the cumulative distribution function of the Poisson law ($\text{LRE}(x,c)\geq 6$). 
In this last, Octave and MatLab presented better results that their previous versions (see~\cite{FreryCilamce2010}). 

When computing the $F$ distributions, Octave produced the best results, mainly if compared with its previous version; as presented by Frery et al.~\cite{FreryCilamce2010}, this platform had produced the worst answers.
But Octave fails to produce acceptable results when dealing with the binomial and t-Student laws. 
Regarding the normal distribution, MatLab and Octave obtained the same good results, while Scilab produced bad results. 
The three platforms were acceptable when dealing with the gamma law, that is, in this case $\text{LRE}(x,c)\geq 6$.


Matlab and Octave failed at computing the t-Student distribution; in every assessed case, there was no match or they returned an error message.
This is a serious issue due to the widely spread use of this distribution in statistical tests.

Six out of eleven linear regression datasets were not adequately dealt with by any of the considered platforms.
Only Matlab provided acceptable results for \texttt{Filip},  \texttt{Norris}, \texttt{Wampler1} and for \texttt{Wampler2}.
\texttt{Wampler2} was acceptably treated by Octave under Windows and Linux  and Scilab under the three operational systems. 
Again, no single platform can be advised as safe for the linear regression problems here considered.

Suprisingly, the same results were provided by the three platforms when making decisions about the determinant of ill-conditioned matrices under the three operating systems. 
The number of erroneus result was acceptable, that is only $94$ in $240$ logical comparisons with the value zero. 
Nevertheless, users are advised to be very careful when testing equality between a value of interest and a numerical computation involving determinats in these platforms.

The assessment based on spectral graph analysis presented a very consistent behavior with respect to the problem size (the bigger the graph, the worse the answer), being $\ell_M$ and $\ell_N$ the most sensitive quantities across all platforms and operating systems, and they can be reported as good in most cases.
The first and last eigenvalues ($\ell_1$ and $\ell_{m+n}$) are always dependable if computed in double precision and then tested in single precision, being the latter consistently more precise than the former.
The balance of bipartite connected graphs did not have a strong impact on the results, except for the percentage of correct eigenvalues.

Extreme care must be taken when making decisions about graphs based on their spectral properties.
As a rule of the thumb, double-precision computation is advised, but the comparison to known values should be made rounding or, at most, using at most floating point representation.

Regarding the variability among operating systems, MatLab and Octave were equivalent and more consistent than Scilab in most of the situations under assessment.

The results are the same in platforms under 32 and 64 bits operating systems, so the latter were not reported in the tables.

\end{document}